\documentclass[twocolumn,showpacs,preprintnumbers,amsmath,amssymb]{revtex4}
 \usepackage{graphicx}
 \usepackage{dcolumn}
 \usepackage{bm}

\begin{document}

\title{Pathway of D$^{+}$ in Sequential Double Ionization of D$_2$ in an Intense Laser Pulse}

 \author{Mohsen  Vafaee}
  \email{mo_vafaee@sbu.ac.ir}
   \affiliation{Laser-Plasma Research Institute, Shahid Beheshti University, G. C., Evin, Tehran 19839-63113, Iran\\}
 \author{Babak Shokri}
  \email{B-Shokri@sbu.ac.ir}
   \affiliation{Laser-Plasma Research Institute, Shahid Beheshti University, G. C., Evin, Tehran 19839-63113, Iran\\}

\newcommand{\be}{\begin{equation}}
\newcommand{\ee}{\end{equation}}
\newcommand{\bea}{\begin{eqnarray}}
\newcommand{\eea}{\end{eqnarray}}
\newcommand{\h}{\hspace{0.30 cm}}
\newcommand{\vs}{\vspace{0.30 cm}}
\newcommand{\n}{\nonumber}
\begin{abstract}
We show the details of the pathway for dissociative ionization process of ground electronic state of aligned D$^{+}_{2}$ due to first ionization of D$_2$ in short ($\thicksim$100 fs) and intense ($4.0\times10^{14}$ W cm$^{-2}$) 480 nm laser pulses. The initial vibrational state of D$^{+}_{2}$ comes from the vertical transformation of the ground state of D$_{2}$.
The initial wavepacket in the ground electronic state of D$_{2}^{+}$ is outgoing through dissociation-ionization channel accompanied by a strong coupling between $1s \sigma_{g}$ and $2p \sigma_{u}$ electronic states. 
We show explicitly that the transition from the coupling states $1s \sigma_{g}$ and $2p \sigma_{u}$ to the ionization state is not a direct transition but takes place through other intermediate states with some dissociation energy that results in the internuclear distribution of the ionization to move considerably to larger internuclear distances.
\\
\end{abstract}
\maketitle
\section{Introduction}
When molecules are exposed to an intense laser pulse (even for the simplest molecules, i.e. H$_{2}$ and H$^{+}_{2}$), a wealth of fascinating phenomena may be observed. A few examples are: above-threshold or tunneling ionization, charge resonance enhanced ionization, dissociative-ionization, above threshold dissociation, bond softening and hardening, high order harmonic generation, etc. \cite {review}.
An interesting subject in studying of H$_{2}$ (D$_{2}$) in the intense laser pulse is double ionization \cite {Alnaser}. 
The double ionization of H$_{2}$ (D$_{2}$) exposed to an intense laser field can proceed through either non-sequential or sequential mechanisms, depending on the intense laser pulse parameters. 
Sequential double ionization (SDI) is the dominant mechanism at higher intensities and also for circularly polarized fields. 

Another interesting new phenomenon has been observed in a recent experimental study of D$_{2}$ exposed to an intense laser field that revealed a new high-energy band for the kinetic energy release (KER) spectrum at short wavelength in the SDI \cite {Litvin}.
In the experiments of Litvinuk et. al., the linearly polarized laser pulses of different peak intensities with duration of $\thicksim$100 fs were focused into a well-collimated supersonic jet of target D$_{2}$ molecules inside the uniform-electric-field ion imaging spectrometer. 
They found a new high-energy band for the KER spectrum in an 480 nm laser pulse  with peak intensity $4\times10^{14}$ W cm$^{-2}$. 
The emergence of this new band depend on the wavelength, i.e. the energy of this band is decreasing with increasing the wavelength until this band merges with the enhanced ionization band for 800 nm and longer wavelength \cite{Litvin}. The position of this high-energy band seems to be independent of the intensity for the various wavelengths.

\begin{figure}
	\resizebox{90mm}{!}{\includegraphics{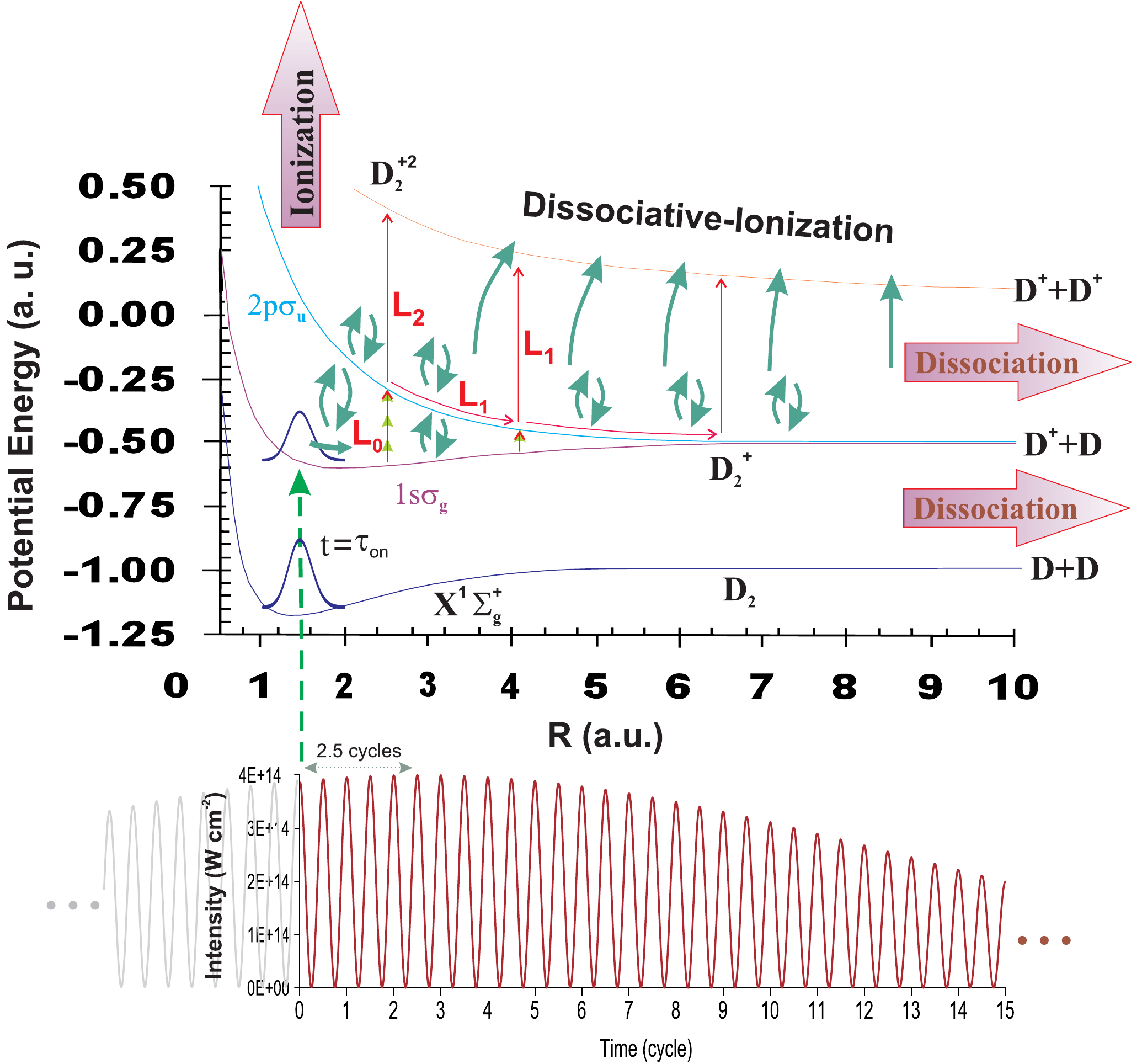}}

	  \caption
		{ 	\label{fig:f1} 
We suppose that at a definite time, $\tau_{on}$, H$_{2}$ (D$_{2}$) is ionized vertically from its ground state to the ground electronic state of H$^{+}_{2}$ (D$^{+}_{2}$). Ionization of H$^{+}_{2}$ (D$^{+}_{2}$) practically occurs through the DIC. The nuclear component in this channel (DIC) posses both dissociation energy (DE) and Coulomb explosion energy (CEE). The green arrows are related to the results of the present simulation.
		}
\end{figure}

To explain this phenomenon, Litvinyuk and et. al. proposed some possible pathways to emerge of this high-band energy (Fig.~\ref{fig:f1}) as follow.
After the neutral molecule is singly ionized, the vibrational wavepacket propagates on the field-dressed electronic potential energy surfaces of D$_{2}^{+}$, staying mostly in the lowest two electronic states (field-free $\sigma_{g}$ and $\sigma_{u}$ states). 
All pathways involve three-photon resonance of the ground state to the first excited electronic state of D$_{2}^{+}$, L$_{0}$.
In one pathway, L$_{1}$ in Fig.~\ref{fig:f1}, the ion's internuclear distance increases in a dissociative state until reaching the critical values ($R_{c}$), where the second ionization takes place with high probability and the Coulomb explosion occurs. The final KER of the nuclear fragments will include the kinetic energy gained during dissociation and also kinetic energy of the Coulomb explosion ($1/R_{c}$).
In another possible pathway, L$_{2}$ in Fig.~\ref{fig:f1}, the second ionization may take place directly at shorter internuclear distances for the shorter wavelengths. In this case the three-photon excitation to the $\sigma_{u}$ state is only a virtual one, with ionization yield being strongly enhanced by the presence of the intermediate resonance. The fragments would gain their full kinetic energy on the Coulomb potential, and that energy will be very close to that of the first discussed pathway, L$_{1}$, since the Coulomb and $\sigma_{u}$ potentials are nearly parallel to each other. The present experimental data are not sufficient to determine which of these two pathways may be the real pathway or near to reality \cite{Litvin}.
Another possible pathway, a net-two-photon pathway (three-photon excitation followed by subsequent de-excitation at the one-photon resonance), would result in lower kinetic energy fragments that is inconsistent with the experimental results \cite{Litvin,Staudte2007PRA}.

To explore the details of the Litvinyuk's experiment and the real pathway, it is necessary to set up a simulation that allows us to declare the details of the dissociation-ionization pathway of D$^{+}_{2}$.  
The dynamics of H$_{2}$ (D$_{2}$) and H$^{+}_{2}$ (D$^{+}_{2}$) exposed to an intense laser field are very complicated because it involves electron and nuclear dynamics simultaneously with different time scales that results in a complex dissociation-ionization process. In intense laser field, electron dynamics occur in attosecond time scales and nuclear dynamics, i.e. vibration and rotation, takes place in femtosecond and picosecond time scales respectively. It is sometimes possible based on Born Oppenheimer approximation (BOA) to investigate the nuclei and electron dynamics, separately. This approach is extensively used to investigate electronic dynamics of molecules in intense laser fields  \cite{zuo1995,peng2003,vafa2006}.
But this assumption fails when both nuclei and electron dynamics are important and considerable. This situation happens with intense $\thicksim$10-100 femtosecond laser pulses, when the duration of the laser pulse is comparable to the vibrational period of molecules \cite{Litvin,pavicic2005,Ergler2005,Rudenko2007,Staudte2007PRL,Staudte2007PRA}.
In these conditions, a complex simulation based on the solving of the time dependent Schr\"{o}dinger equation (TDSE) beyond the BOA would be required for the complete description of molecular dynamics. 
For a molecule with two or more electrons, this task is very far from present available computer ability, even without consideration of the nuclear dynamics. 
The rigorous approach is feasible only for linear molecules aligned with the electric laser field with only one electron. In this condition, it is possible to consider the full dimensional electron dynamics and also the vibrational nuclear dynamics in the solution of the TDSE for H$^{+}_{2}$ (D$^{+}_{2}$) \cite {vafa2008,Hu2009}. 

In this work, we have performed a simulation for D$^{+}_{2}$ beyond the BOA and also without the soft-core approximation \cite {Su-Eberly1991, Saugout2008}, i.e., 
by the rigorous solution of the TDSE for the full dimensional electron dynamics of the aligned D$^{+}_{2}$ with the electric laser field.
This paper is organized as follows: the details of the numerical implementation and the simulation setup are presented in Sec. II. Results and discussions appear in Sec. III. Section IV contains observations and concluding remarks. Throughout this article we use the atomic units unless stated otherwise. 
\section{Numerical solution of the TDSE}
%
%
Time dependent Schr\"{o}dinger equation in the cylindrical polar coordinate system for H$^{+}_{2}$ (D$^{+}_{2}$) located in the laser field parallel to the internuclear axis in atomic unit (au) can be expressed as
\begin{eqnarray}\label{eq:1}
i\frac{\partial \psi(z,\rho,R,t)}{\partial t}=H(z,\rho,R,t) \psi(z,\rho,R,t)
\end{eqnarray}
where the Hamiltonian for this system, $H(z,\rho,R,t)$, is given by \cite{Hiskes,ban-lu2000,vafa2004}
\begin{eqnarray}
H(z,\rho,R,t) =
&&-\frac{2m_{N}+m_{e}}{4m_{N}m_{e}} \left[\frac{\partial ^2}{\partial \rho^2}+\frac{1}{\rho}
 +\frac{\partial }{\partial \rho}+\frac{\partial ^2}{\partial z^2}\right]\nonumber\\
&&-\frac{1}{m_{N}}\frac{\partial ^2}{\partial R^2}+V_{C}(z,\rho,R,t),
\label{eq:2}
\end{eqnarray}
\begin{eqnarray}
V_{C}(z,\rho,R,t) =
&& -\frac{1}{\sqrt{\left( z+\frac{R}{2}\right)^2+\rho^2 }}
  -\frac{1}{\sqrt{\left( z-\frac{R}{2}\right)^2+\rho^2 }}\nonumber\\
&&+ \frac{1}{R}+\left(\frac{2m_p+2m_e}{2m_p+m_e}\right) z E_0 f(t)\cos(wt),
\label{eq:3}.
\end{eqnarray}
with $E_0$ being the laser peak amplitude, $\omega = 2\pi \nu $ the angular frequency, and finally $f(t)$ 
the laser pulse envelope which is set as
\begin{eqnarray}
f(t) =exp \left[ \frac{-2\ln(2)(t-t_{on})^2}{\tau^2_p} \right]
\label{eq:4}
\end{eqnarray}
where $ \tau_p $ is the full width at half maximum (FWHM) duration of the Gaussian shape of the laser pulse, with $ \tau_p $=25 cycles in this work. 
For the time discretization of the TDSE, a propagator derived from split-operator methods has been used. This propagator is unitary and is obtained by combining the classical split operator and the Crank-Nicholson method \cite{splitting}.
The details of our calculations are described in our previous reports \cite{vafa2004,vafa2006,vafa2008}. 
In this simulation, the time step is set to  $\delta t=0.02$ and $\delta t_{R}=0.2$ for the electron  and nuclei time propagation respectively and $\tau_p $=25 cycles ($\thicksim$40 fs). 
The differential operators in Eq. (\ref{eq:2}) are discretized by the eleven-point difference formulae which have tenth-order accuracies \cite{vafa2006}. 
To solve the above TDSE numerically, we adopted a general non-linear coordinate transformation for both electronic and nuclear coordinates. 
For the spatial discretization, we have constructed a finite difference scheme with a non-uniform (adaptive) grid for $z$ and $\rho$ electronic coordinates which are finest near the nuclei and coarsest at the border regions of the simulation box. A finite difference scheme with an adaptive grid is used also for $R$ coordinates that is finest for small $R$ and become a coarse grid for large $R$. 
Using a fine grid for electronic coordinate ($z$ and $\rho$) near the nuclei and for small values of internuclear distance coordinate ($R$) improves the treatment of the electron dynamics near the nuclei (the Coulomb singularities) and bound states of the nuclear dynamics, while the use of a coarse grid near the borders improves the speed of calculations. 
The grid points for $z$, $\rho$, and $R$ coordinates in our simulation are 560, 155, and 240 respectively. The finest grid sizes values in this adaptive grid schemes are 0.13, 0.2, and 0.025 respectively for $z$, $\rho$, and $R$ coordinates. The grids extend up to $z^{max}=54$, $\rho^{max}=25$, and $R^{max}=16$. 

The simulation setup is shown schematically in Fig.~\ref{fig:f1}.
The laser pulse in this simulation suddenly turns on at time $\tau_{on}$. We are assuming $\tau_{on}$ is the time on which a H$^{+}_{2}$ (D$^{+}_{2}$) is suddenly created according a Frank-Condon transition from the neutral H$_{2}$ (D$_{2}$) to H$^{+}_{2}$ (D$^{+}_{2}$) $\sigma_g $ state (Fig.~\ref{fig:f1}). 
In an intense laser field, some parts of D$^{+}_{2}$ wavepackage become unbound and outgoing through different channels. A part of this unbound wavepacket becomes outgoing as $D^{+}+D$ through dissociation channel (DC) and another part becomes outgoing through dissociation-ionization channel (DIC) as $D^{+}+D^{+}$ (Fig.~\ref{fig:f1}). 
The nuclear components in the dissociation-ionization channel possess both dissociation and Coulomb explosion energies \cite{vafa2008}.

\section{Results and Discussion}
\begin{figure*}
		\resizebox{120mm}{!}{\includegraphics{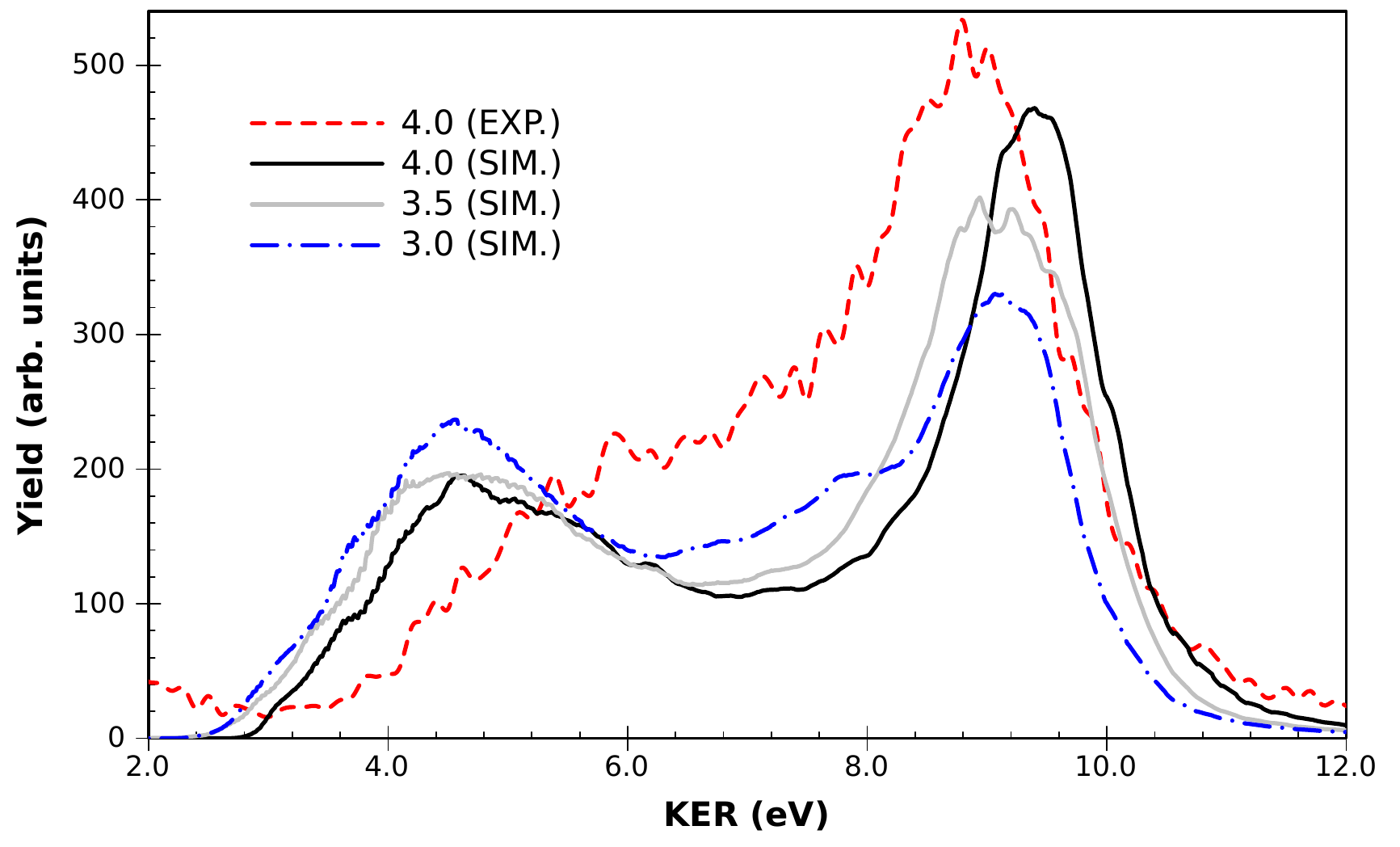}}
	\caption
		{ \label{fig:cKER}
Comparison of the experimental and simulation KER spectra. The  red dotted line is related to the experimental results \cite{Litvin} for a 480 nm laser pulse with peak intensity I=$4.0\times10^{14}$ W cm$^{-2}$ and $ \thicksim $100$ \pm 10 $ fs pulse duration. The black solid, grey solid, and blue dot-dash lines show simulation results for 480 nm, $\tau_p\simeq$40 fs laser pulses with peak intensities $4.0\times10^{14}$, $3.5\times10^{14}$  ,and $3.0\times10^{14}$ W cm$^{-2}$ respectively. 
		}
\end{figure*}

\begin{figure*}
\resizebox{120mm}{!}{\includegraphics{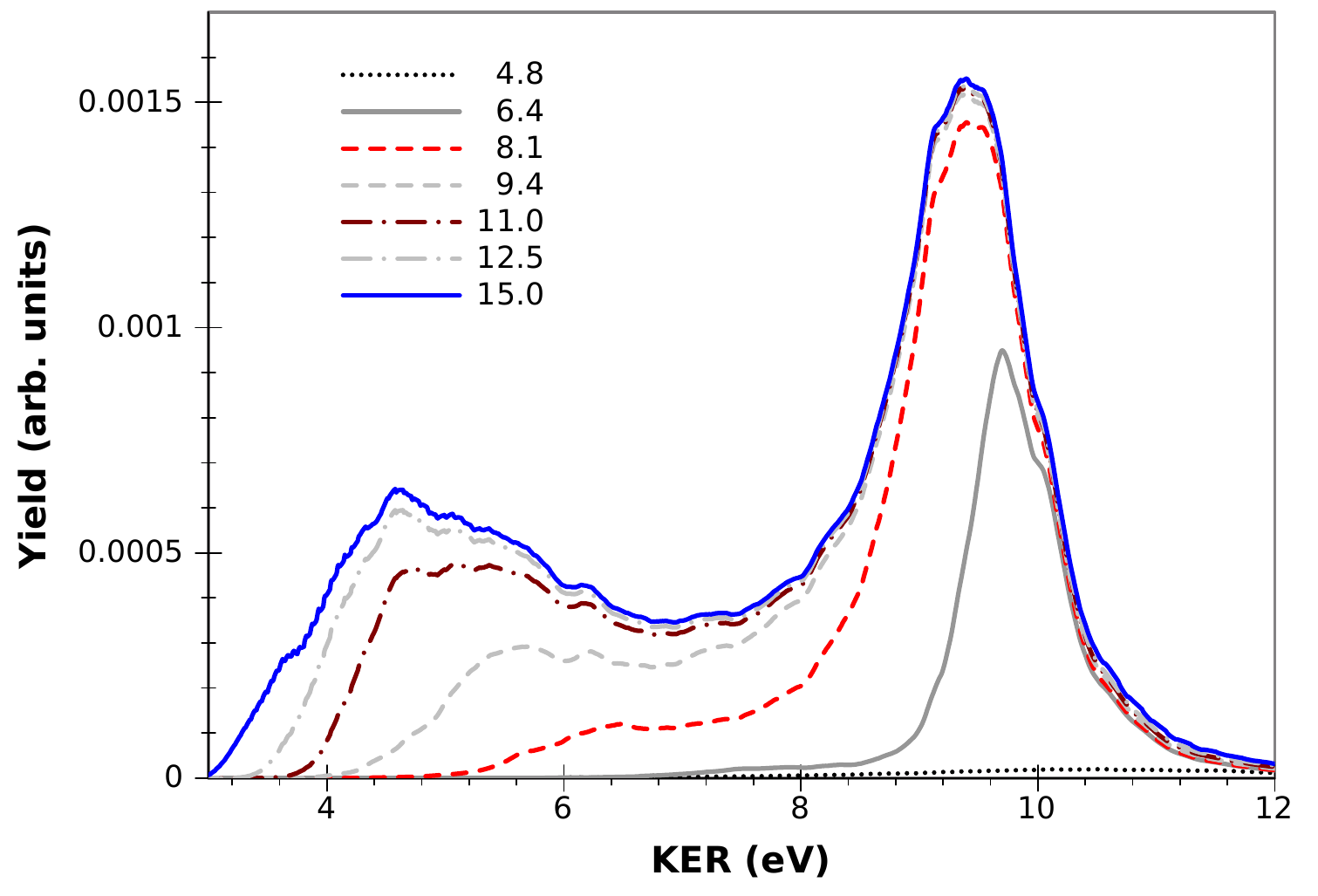}}
\caption{\label{fig:tKER}
Time-dependent construction of the KER Spectrum that is related to the peak intensity $4.0\times10^{14}$ W cm$^{-2}$ shown in Fig.~\ref{fig:cKER}.
The high-energy band of the KER Spectrum appears at first between $\thicksim$4.8 to $\thicksim$8 cycles of the laser pulse. After this time, the low-energy band of the KER Spectrum appears.
		}
\end{figure*}

Figure~\ref{fig:cKER} shows the simulation KER spectra and compares with the experimental result. In this figure, the dotted red line is related to the experimental result \cite{Litvin} for the 480 nm linearly polarized laser pulse with  I=$4.0\times10^{14}$ W cm$^{-2}$ intensity and $\thicksim$100 fs pulse duration. The black solid, grey solid, and blue dot-dash lines show simulation results that are respectively related to $4.0\times10^{14}$, $3.5\times10^{14}$, and $3.0\times10^{14}$ W cm$^{-2}$ intensities and a 480 nm wavelength. The simulation results show explicitly two bands, a high-energy and a lo-energy band which also have appeared in the experimental results \cite{Litvin}.
Some inconsistencies can be seen in Fig.~\ref{fig:cKER} between the experimental and simulation KER spectra related to I=$4.0\times10^{14}$ W cm$^{-2}$ intensity. These inconsistencies are due to some experimental uncertainties in the shape and duration of the laser pulse and the unavoidable simplifications and approximations applied in the simulation setup.
However, Fig.~\ref{fig:cKER} shows many similarities and correspondence between experimental and simulation results. 
The simulation results, in correspondence to the experimental results \cite{Litvin}, show that the decreasing of the intensity of the laser pulse causes a decrease in the signal of the high-energy band with respect to the low-energy band (Fig.~\ref{fig:cKER}).
To derive the details of the KER structure, we represent a time-dependent construction of this signal in Fig.~\ref{fig:tKER}.
As can be seen in this figure, at first the high-energy band appears between $\thicksim$4.8 to $\thicksim$8.1 cycles of the laser pulse 
(note that the meaning of $t=0$ in our simulation, as shows in Fig.~\ref{fig:f1}, is 2.5 cycles earlier than of the peak of the laser pulse).
After this duration, it seems that the pathway of the ionization with high-energy fragment is closed and another pathway of the ionization with low-energy fragment is opened. Therefore, this figure proposes that there are two distinct pathways for the ionization with different energies and time domains.

\begin{figure*}
  \begin{center}
    \begin{tabular}{cc}
      \resizebox{80mm}{!}{\includegraphics{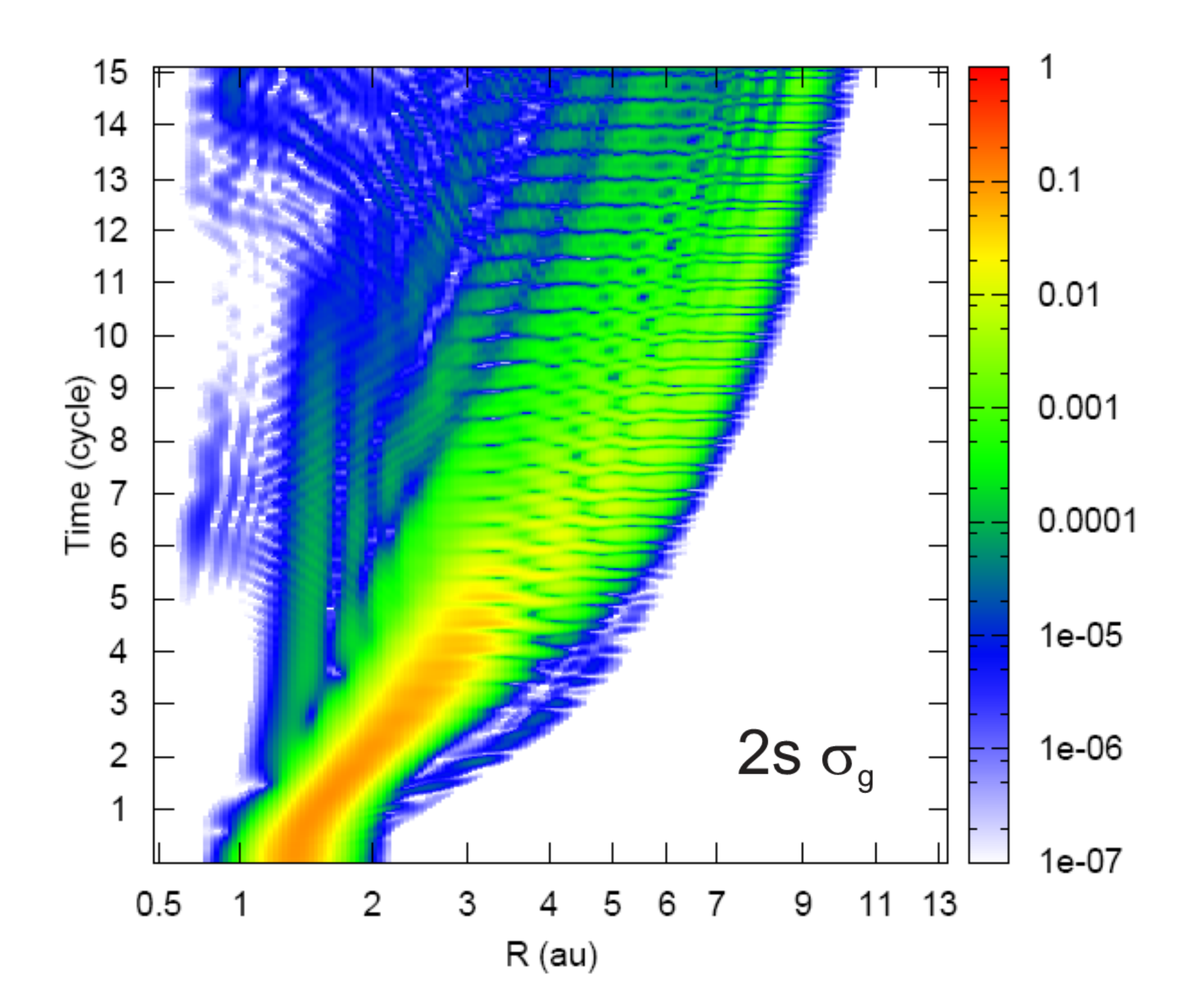}} &
      \resizebox{80mm}{!}{\includegraphics{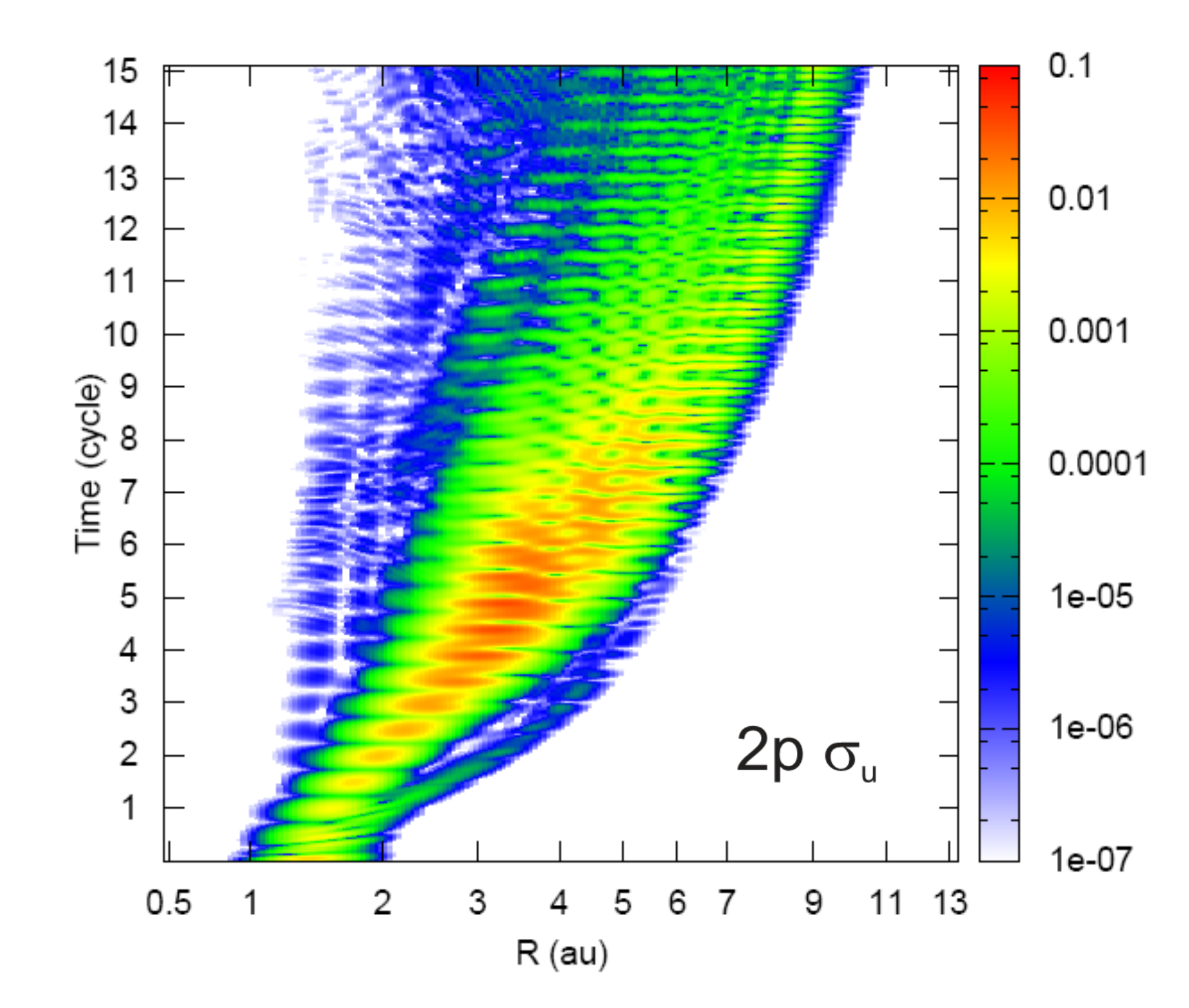}} \\
      \resizebox{80mm}{!}{\includegraphics{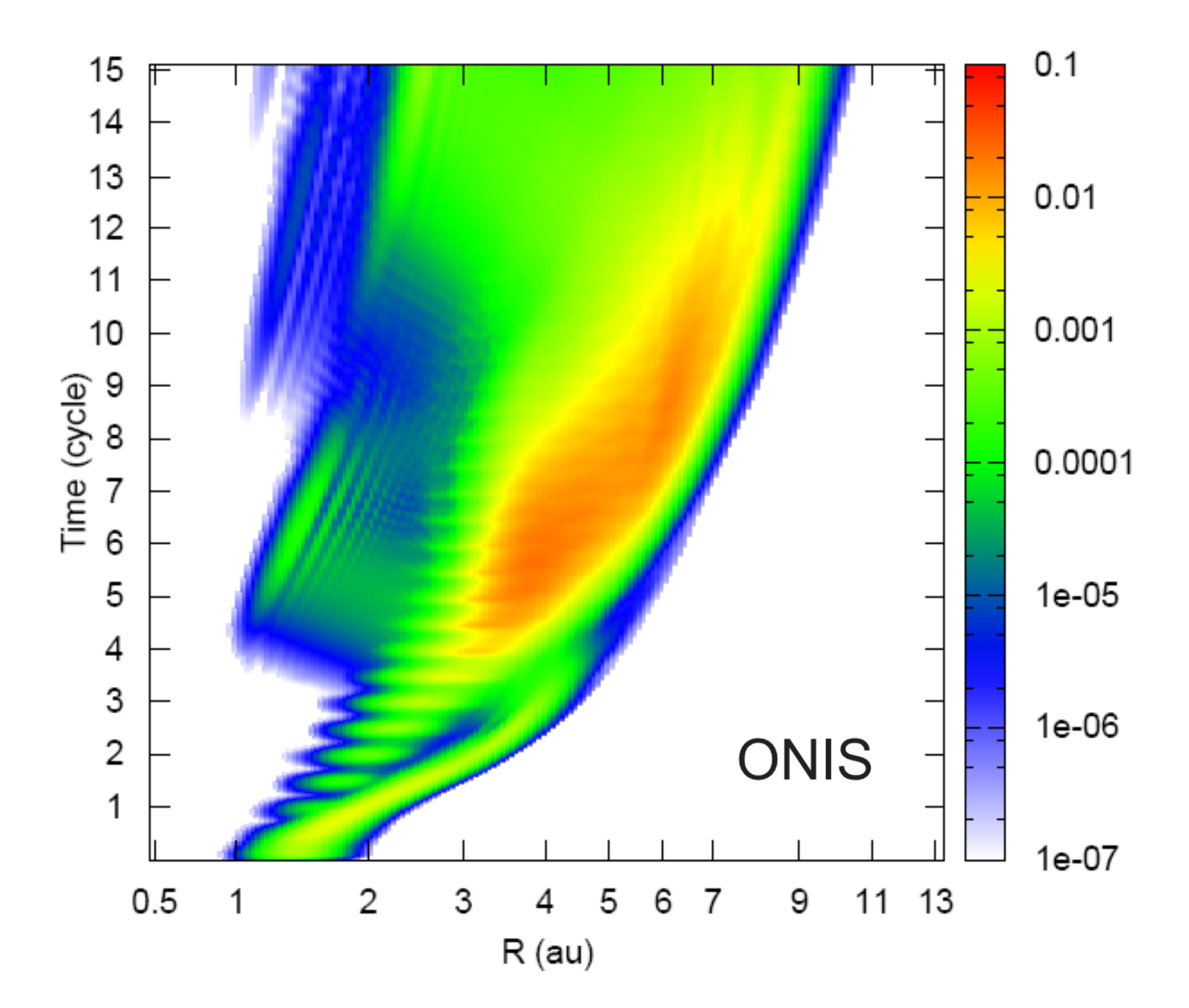}} &
      \resizebox{80mm}{!}{\includegraphics{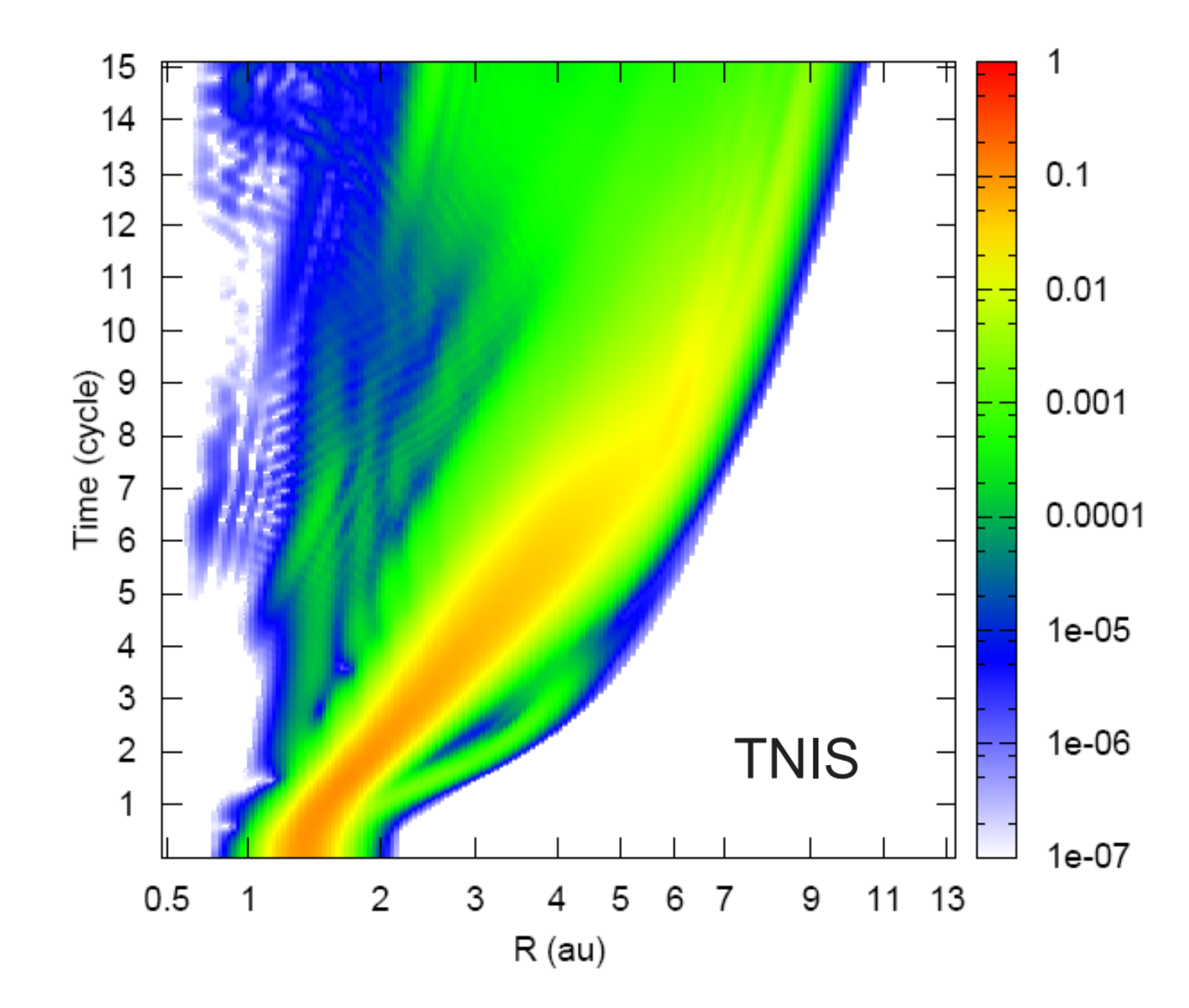}} \\
    \end{tabular}
  \end{center}
  \caption
  { \label{fig:TSP}  
Space-time population of the ground state ($1s \sigma_{g}$), first exited state ($2p \sigma_{u}$), other non-ionized states (ONIS), and total non-ionized state (TNIS). In this figure, R shows the internuclear distance of D$^{+}_{2}$. 
  }
\end{figure*}

Figure~\ref{fig:TSP} represents the space-time population of the ground state ($1s \sigma_{g}$), first exited state ($2p \sigma_{u}$), other non-ionization states (ONIS), and total non-ionization states (TNIS). The difference between the population of TNIS and the population of $1s \sigma_{g} + 2p \sigma_{u}$ equals to the space-time population of the ONIS. 
This Figure shows explicitly the charge resonance among $1s \sigma_{g}$, $2p \sigma_{u}$, and ONIS. The resonance between $1s \sigma_{g}$ and $2p \sigma_{u}$ is very strong and a weaker coupling of these two states with the ONIS are visible in the ONIS's figure. 
These figures show explicitly the space-time dissociation-ionization pathways of D$^{+}_{2}$. Note that, the reduction of the population of the TNIS in the dissociation pathways is due to the ionization. 
Figure~\ref{fig:TSP} shows that the population of $2p \sigma_{u}$ becomes considerable over $\thicksim$2-5 internuclear distances and $\thicksim$3-8 cycles and for ONIS over $\thicksim$3-7 internuclear distances and $\thicksim$5-11 cycles. 
D$^{+}_{2}$ in the dissociative pathways is considerably ionized and will has little population before it dissociates, as has shown in the next figures. 

\begin{figure*}
\resizebox{120mm}{!}{\includegraphics{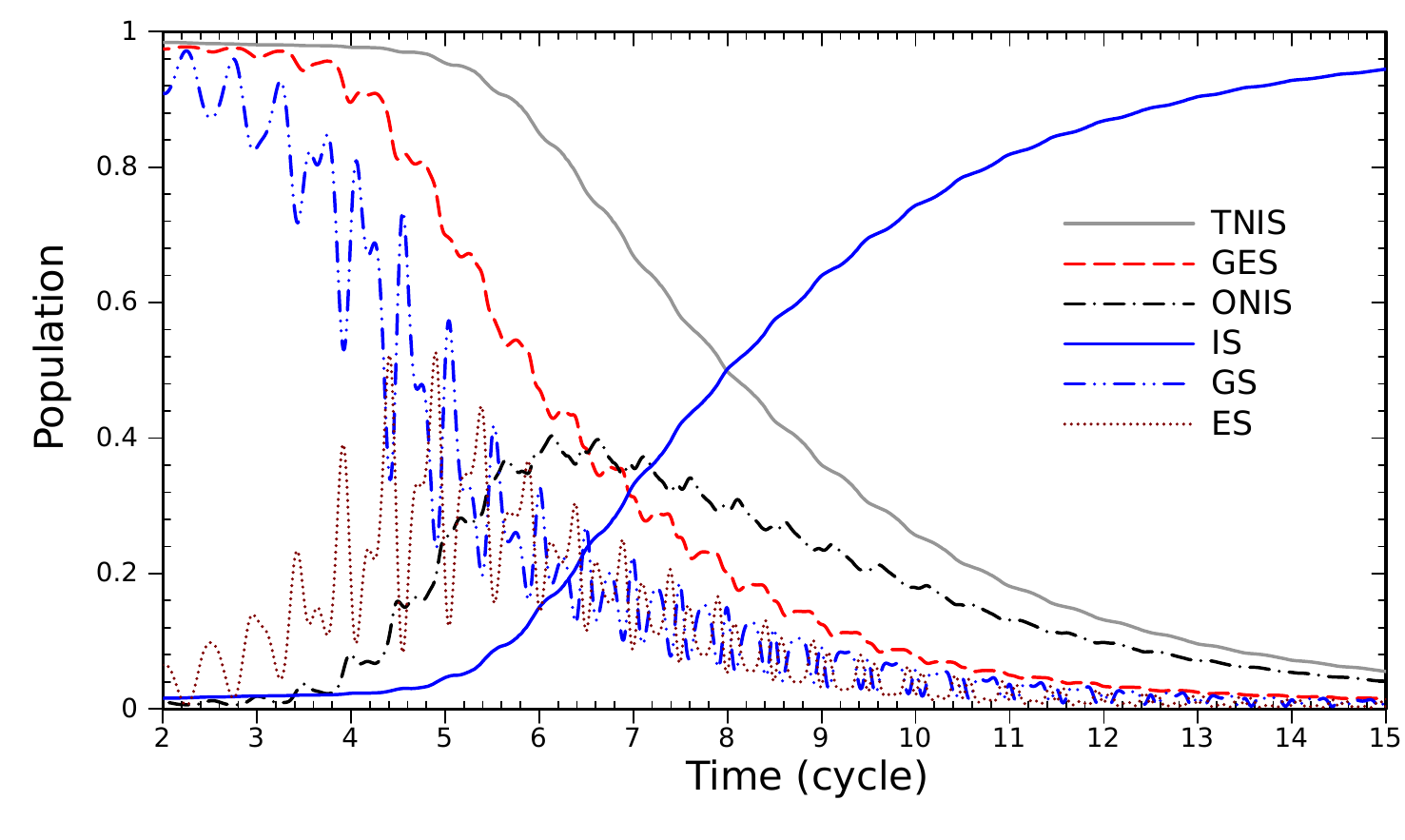}}
\caption{ \label{fig:norm}
Time-dependent population of different states of D$^{+}_{2}$ related to the $4.0\times10^{14}$ W cm$^{-2}$ intensity in Fig.~\ref{fig:cKER}.		
		}
\end{figure*}

To see the details of the dissociation ionization pathways, it is important to derive separately the space-dependent population and time-dependent population of different states. 
Fig.~\ref{fig:norm} represents the time-dependent population of the ground state ($1s \sigma_{g}$, GS), first exited state ($2p \sigma_{u}$, ES), the ground and first exited states together (GES), other non-ionized states (ONIS), total non-ionized states (TNIS) or norm, and finally the time-dependent population of the ionized states (IS). 
As can be seen in this figure, the population of the ONIS increases considerably from about 4 cycles. The charge resonance between different states is observable. A strong charge resonance between Gs and Es appears after 2 cycles results in the population of $2p \sigma_{u}$ becoming more than that of $1s \sigma_{g}$ state occasionally.
In this figure a weaker charge resonance is also observed between the GES and ONIS.    

\begin{figure*}
\resizebox{120mm}{70mm}{\includegraphics{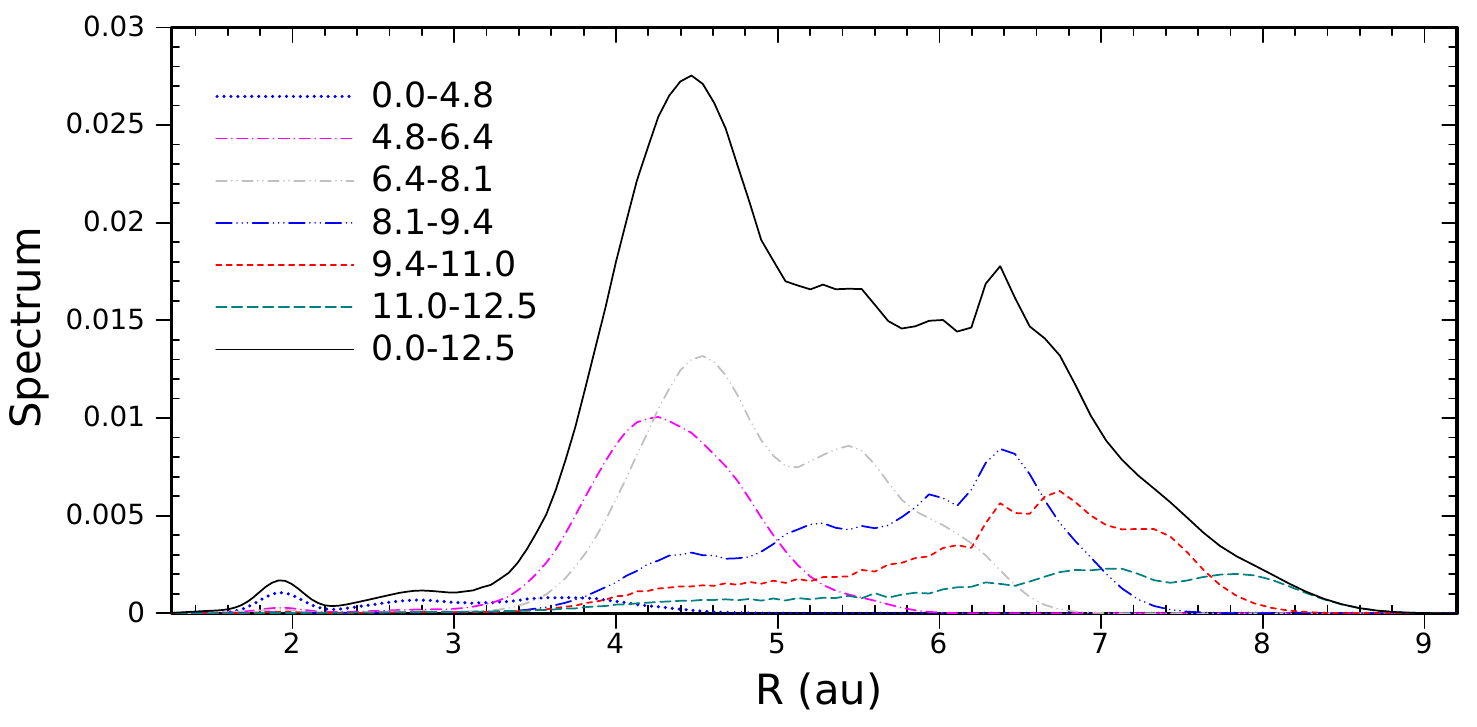}}
\caption{ \label{fig:total_IR}
The Internuclear population of the ionization of D$^{+}_{2}$ related to the $4.0\times10^{14}$ W cm$^{-2}$ intensity in Fig.~\ref{fig:cKER}.
		}
\end{figure*}

Figure~\ref{fig:total_IR} shows the internuclear populations of the ionization of D$^{+}_{2}$ related to the $4.0\times10^{14}$ W cm$^{-2}$ intensity in Fig.~\ref{fig:cKER}. 
The internuclear populations of the ionization of D$^{+}_{2}$ in Fig.~\ref{fig:total_IR} have been derived by the virtual detector method \cite{Feuerstein2003JPB,vafa2004}.
Indeed, this figure represents the R-dependent ionization rate, with an arbitrary unit, of D$^{+}_{2}$ beyond BOA \cite{vafa2008}.
The BOA approach has often been used to derive the R-dependent ionization rate of D$^{+}_{2}$ in most previous studies \cite{zuo1995,peng2003,sabz2005}. 
In this figure, we also show the internuclear populations for the different time intervals shown in Fig.~\ref{fig:tKER}. This figure shows explicitly that the high-energy band of the KER spectrum in Fig.~\ref{fig:tKER}, that is constructed between $\thicksim4.8-8$ cycles, has a strong ionization rate about $\thicksim4-5$ internuclear distances and the low-energy band of the KER spectrum (Fig.~\ref{fig:tKER}), that is constructed between $\thicksim8.1-12.5$ cycles, has a relatively strong ionization rate about $\thicksim6-7$ internuclear distances. After 12.5 cycles, the intensity of the laser field becomes considerably low (Fig.~\ref{fig:f1}) and a significant population of D$^{+}_{2}$ was becomes ionized as shown in Fig.~\ref{fig:norm}. Therefore it is impossible to appear an enhanced ionization signal for larger internuclear distances.

\begin{figure*}
  \begin{center}
    \begin{tabular}{cc}
      \resizebox{75mm}{45mm}{\includegraphics{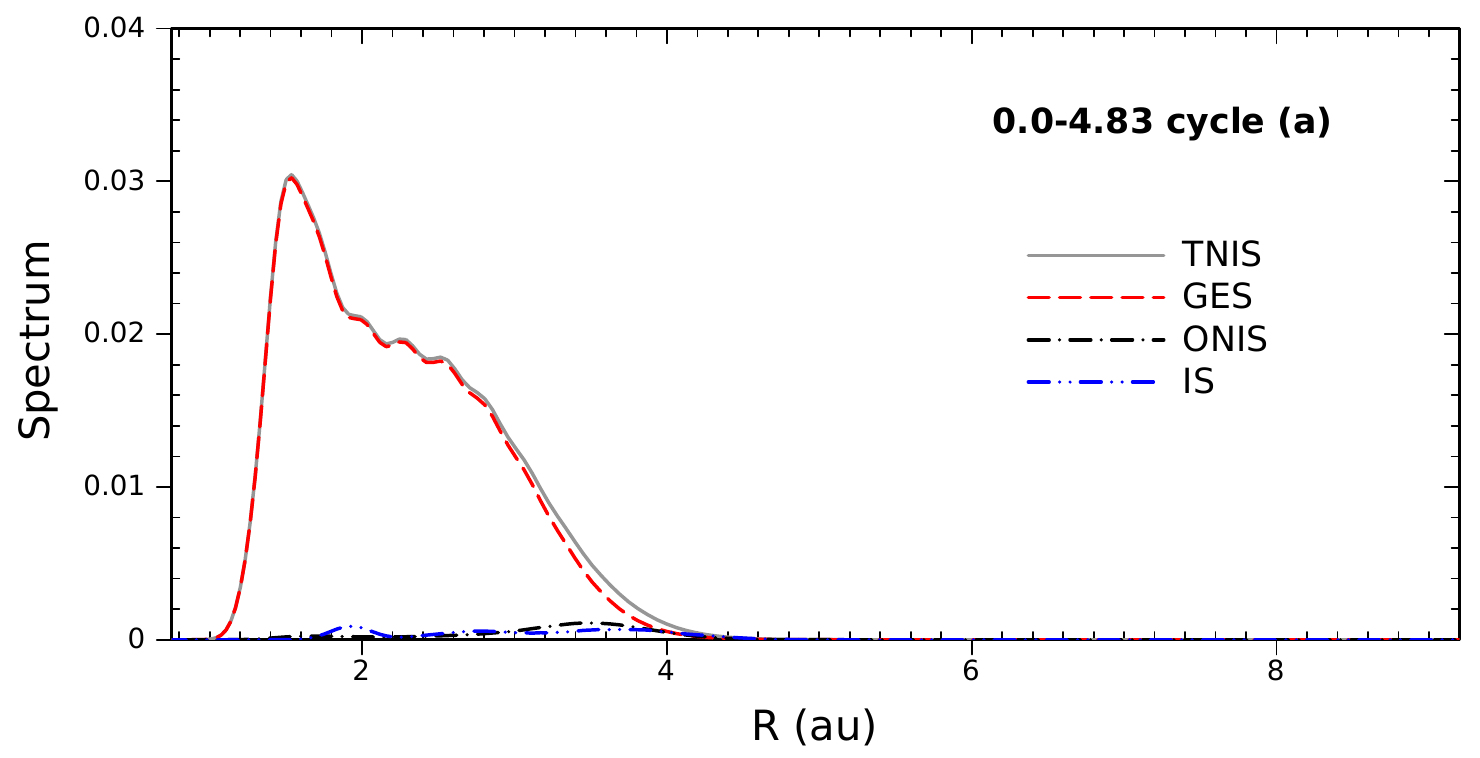}} &
      \resizebox{75mm}{45mm}{\includegraphics{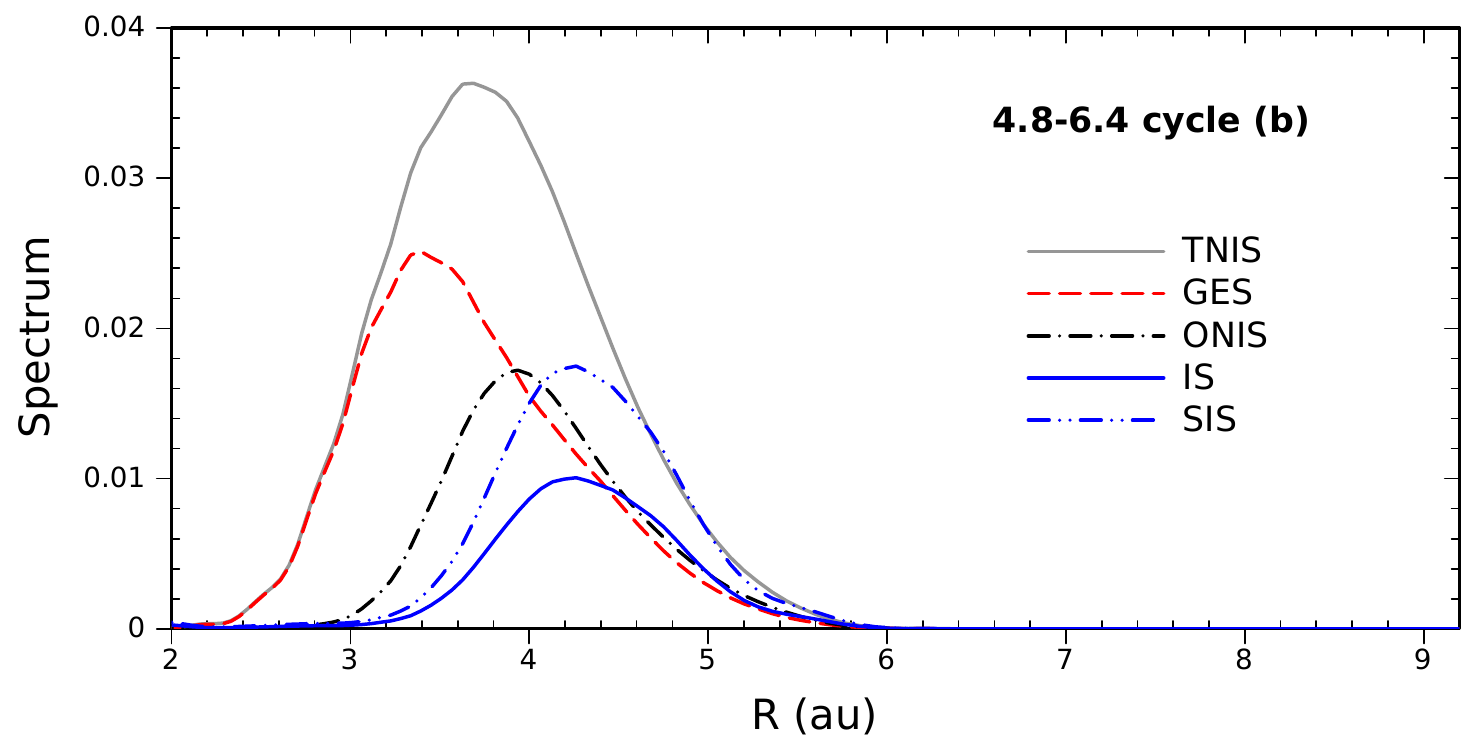}} \\
      \resizebox{75mm}{45mm}{\includegraphics{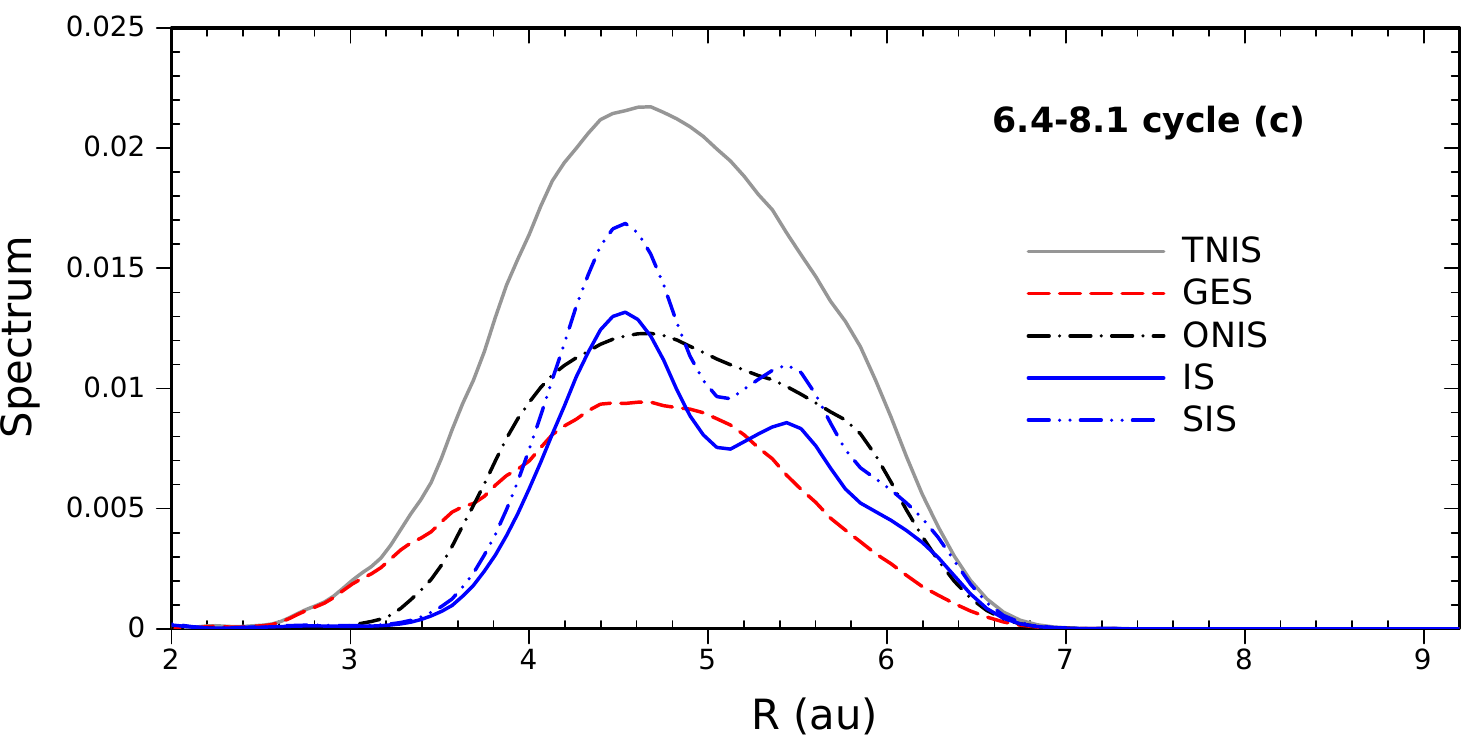}} &
      \resizebox{75mm}{45mm}{\includegraphics{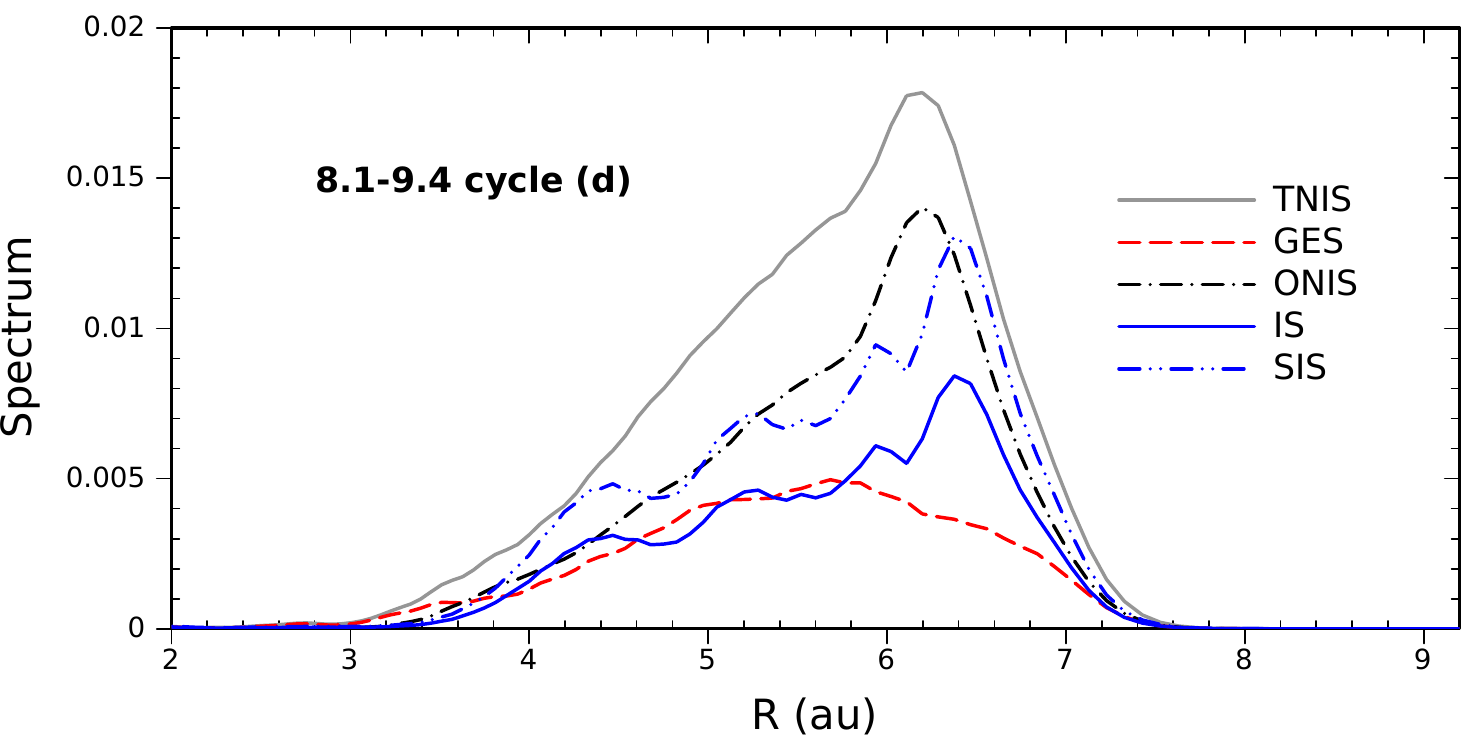}} \\
      \resizebox{75mm}{45mm}{\includegraphics{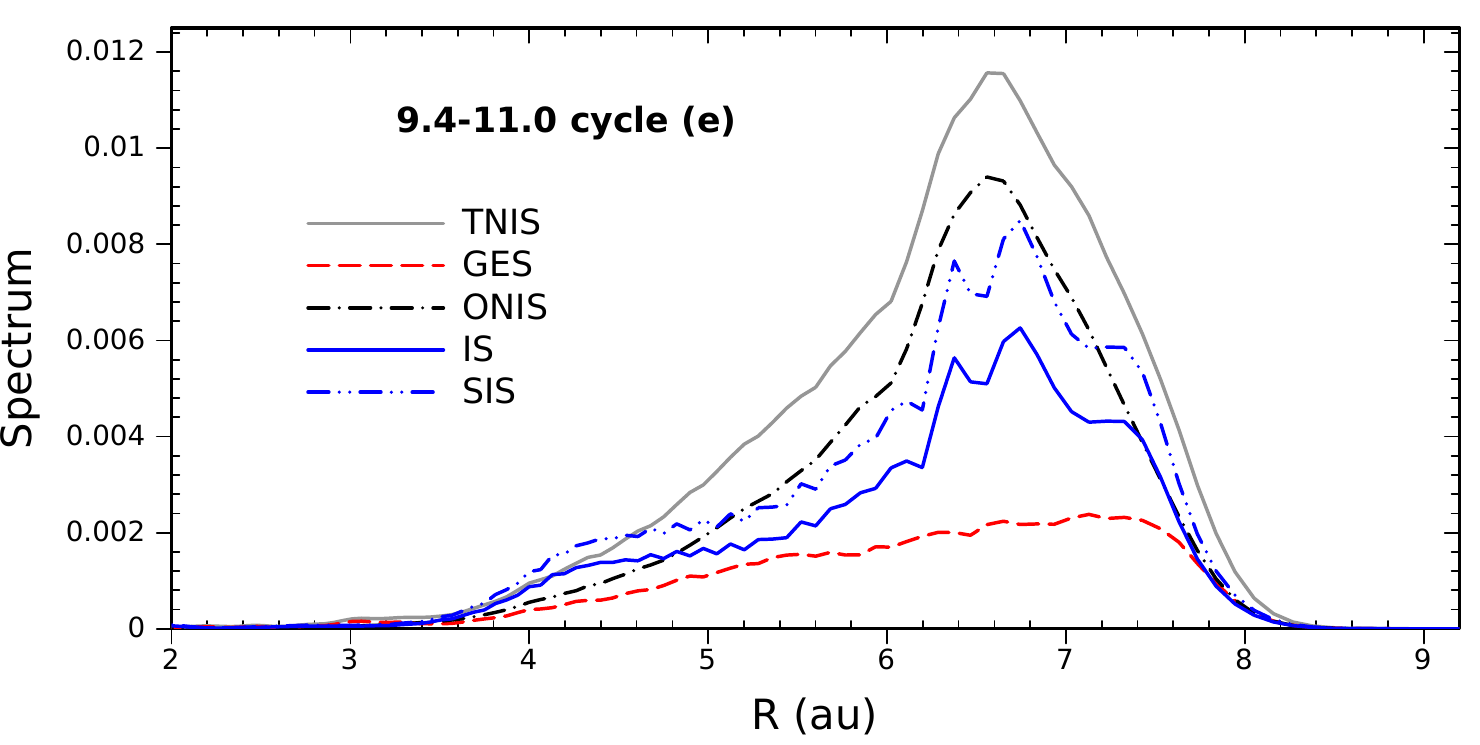}} &
      \resizebox{75mm}{45mm}{\includegraphics{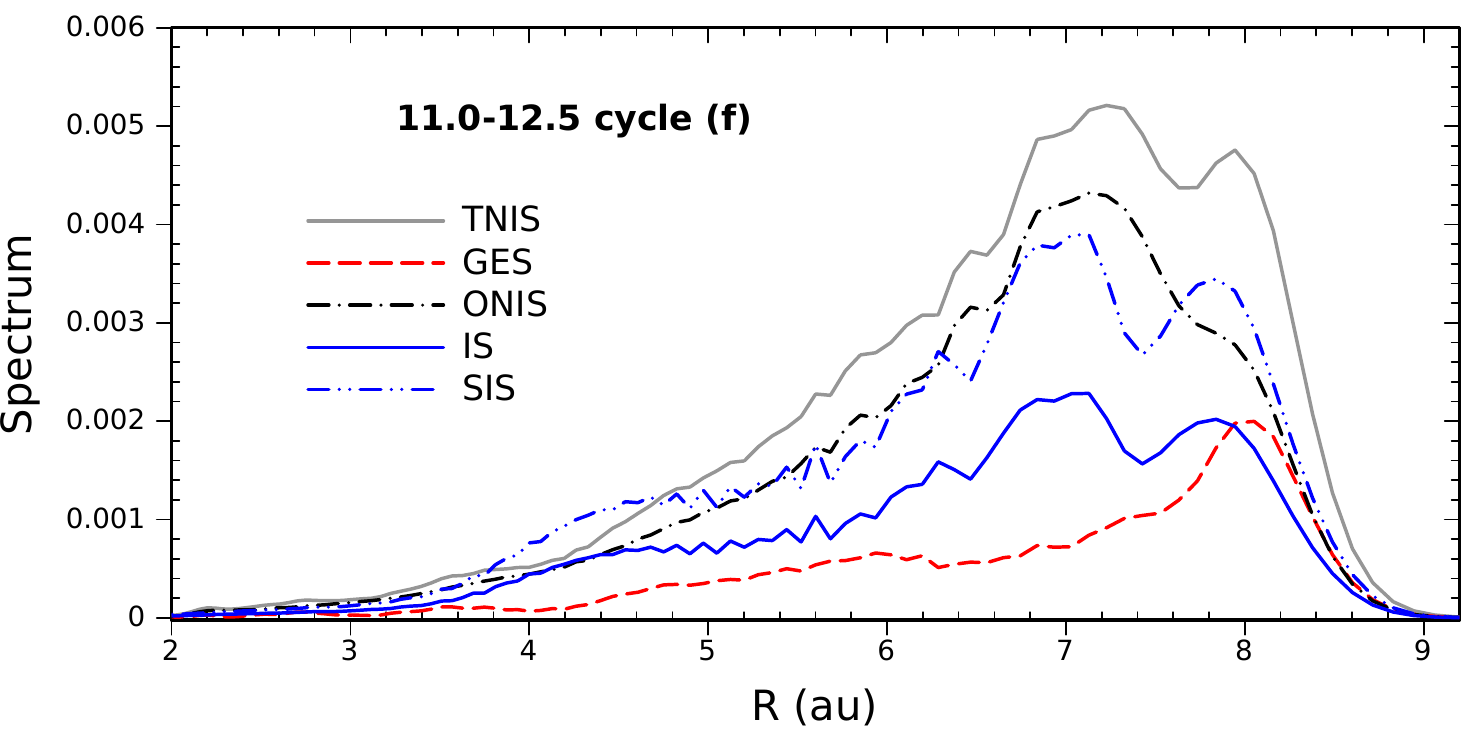}} \\
    \end{tabular}
  \end{center}
  \caption
  { \label{fig:RKER}
IS (solid blue lines) shows outgoing internuclear population from the dissociation ionization channel of D$_{2}^{+}$ for the specified time interval on each figure. Other lines as follows: the time-average internuclear population of the TNIS (solid grey lines), GES (dashed red lines) and ONIS (dot-dash black lines). To better comparison of the IS and ONIS values, the IS overall population are scaled to the ONIS overall population and showed by the dot-dot-dash blue lines (SIS). 
  }
\end{figure*}

In Coulomb explosion imaging \cite{cloumbimaging}, we want to relate the KER spectra to the internuclear population of the wavepacket of molecules and so derive the time-dependent structure of the molecules. Figures~\ref{fig:tKER} and ~\ref{fig:total_IR} can relate the structure of the KER spectrum and the structure of the internuclear population of the ionization. If we show the relation of the time-dependent internuclear population of the non-ionized D$^{+}_{2}$ wavepacket and the time-dependent internuclear populations of the ionized D$^{+}_{2}$, it clearly displays the relation of the time-dependent structure of molecules and the KER spectrum.
In Fig.~\ref{fig:RKER}, we present the time-averaged internuclear population of different states of D$^{+}_{2}$ for the same time intervals that is indicated in Fig.~\ref{fig:total_IR}. The different states in Fig.~\ref{fig:RKER} are chosen as follow: the TNIS (solid grey lines), the GES (dashed red lines), and the ONIS (dot-dash black lines). In this figure, for better comparison, we also show the internuclear populations of the ionization in Fig.~\ref{fig:total_IR} as IS (ionized states) by solid blue lines.
Among the different states of D$_{2}^{+}$ in Fig.~\ref{fig:RKER}, it is obvious that the ONIS is very similar to the IS.  
For the better comparison of the IS and ONIS spectra, in each figure we have scaled the overall population of the IS to the overall population of the ONIS and presented as scaled ionized state (SIS, dot-dot-dash blue lines) in Fig.~\ref{fig:RKER}. This figure shows that the structure of the SIS is very similar to the structure of ONIS and therefore we can conclude that the origin of the KER spectrum is from the ONIS and the KER spectrum can relatively map the time-dependent structure of the ONIS. 

Figure~\ref{fig:tKER} to \ref{fig:RKER} constitute a complete set that enable us to trace the detail events occurring during dissociative ionization of D$_{2}^{+}$ which can be expressed as follows.
During the 0 to $ \thicksim5$ cycles, Fig.~\ref{fig:norm} shows a strong coupling between GS and ES and the GS is outgoing through a DC (Fig.~\ref{fig:TSP}). In this duration, Fig.~\ref{fig:norm} shows that the majority of the population of the TNIS is in the GES and the populations of the other states are negligible. Near the end of this time interval, the population of the ONIS becomes considerable.   
Among different populations in Fig.~\ref{fig:RKER}(b), the IS (SIS) and ONIS are very similar but there is a little displacement in the population of the IS with respect to the TNIS.
This figure also shows that 
the structure of the IS and the GES are similar but the IS shifts to right, i.e., to larger internuclear distances with respect to the GES. This displacement shows a considerable DE for the IS. 
Fig.~\ref{fig:RKER}(b) also shows that the nuclear fragments of the IS have considerable CEE because the IS distribution has small internuclear distance distribution.

During the ~5-8 cycles of the laser pulse, the high-energy band of the KER spectrum is constructed (Fig.~\ref{fig:tKER}). Fig.~\ref{fig:RKER}(b-c) show the internuclear populations of the different states for this band. Fig.~\ref{fig:RKER}(b-c) also show that among different states, the ONIS has the closest structure to IS (SIS). 
Therefore, this confirms that the IS does not originate directly from the GES but it arises from the ONIS. 
This point is represented schematically in Fig.~\ref{fig:f1}. 
Generally, all sections in Fig.~\ref{fig:RKER} approve that the structure of the IS is very similar to the ONIS, for low or high-energy band of KER spectrum.  

Comparing of Fig.~\ref{fig:RKER}(a-c) reveals that during ionization, different states are outgoing through the DC because their internuclear populations shift to larger internuclear values. 
In Fig.~\ref{fig:RKER}, we can see that from 4.8 to 12.5 cycles, the magnitude of the displacement of the IS with respect to the GES and ONIS is reduced, i.e. the DE of the IS is reduced gradually. We can also see that the peak of the IS spectrum is placed on the right of the GES spectrum in Fig.~\ref{fig:RKER}(b) opposite to that in Fig.~\ref{fig:RKER}(f). Therefore, Fig.~\ref{fig:RKER} shows explicitly that the transition from the GES to the IS is not a vertical transition and the population transfer from the GES to the IS occurs through the intermediate states (ONIS). 
Correspondence between the IS (SIS) and ONIS spectra in Fig.~\ref{fig:RKER}(d-f) are very good and the displacement between the GES and the IS (or ONIS and the IS) becomes less with respect to Fig.~\ref{fig:RKER}(b-c). Therefore, for these space-time intervals that the low-energy band of the KER spectrum is constructed, i.e. Fig.~\ref{fig:RKER}(d-f),during the transition of the population from the GES to the IS, the ionized fragments obtain  the smaller DE and their energy come mainly from the CEE.
\section{Conclusion}

In summary, the pathway for dissociative ionization process of the aligned ground electronic state of D$^{+}_{2}$ exposed to a short ($\thicksim$100 fs) and intense ($4.0\times10^{14}$ W cm$^{-2}$) 480 nm laser pulse that its initial vibrational state comes from vertical transformation of the ground state of D$_{2}$ can be expressed as follows.
During the $\thicksim$5 cycles of the laser pulse, the ground state of D$_{2}$ is strongly coupled with $2p \sigma_{u}$ (Fig.~\ref{fig:f1} and Fig.~\ref{fig:norm}). 
Within 3 cycles, from $\thicksim$5 to $\thicksim$8 cycles of the laser pulse, the high-energy band of the KER spectra is constructed (Fig.~\ref{fig:tKER}). During this time period, a considerable population of the GES transfers to the ONIS and then to the IS (Fig.~\ref{fig:norm}, Fig.~\ref{fig:total_IR}, and Fig.~\ref{fig:RKER}(b-c)). 
The ionized fragments related to this high-energy band is originated mainly from the $\thicksim4-5$ internuclear distances (Fig.~\ref{fig:TSP}, Fig.~\ref{fig:total_IR}, and Fig.~\ref{fig:RKER}(b-c)) and their internuclear distributions (SIS) are very similar to the internuclear distributions of the ONIS, but with some small displacements to larger internuclear distances and also there is some deformation with respect to the ONIS (Fig.~\ref{fig:RKER}(b-c)). 
Therefore, the structure of the IS in Fig.~\ref{fig:RKER} does not map the GES but it is similar to the structure of the ONIS. In another words, the transition from the GES to the IS is not direct but this occurs through the intermediate states (ONIS) with a considerable DE that results in the internuclear distribution of the ionization moves considerably to larger internuclear distances. Therefore, an R-dependent displacement between the IS, ONIS, and the GES structures are observed. 

After $\thicksim$8 cycles of the laser pulse, the laser field is still strong enough and also the population of the TNIS is still significant. 
The TNIS (i.e. GES plus ONIS) continues outgoing through the DIC (Fig.~\ref{fig:TSP}) and reaches a large critical internuclear distance (Fig.~\ref{fig:TSP} , Fig.~\ref{fig:total_IR} ,and Fig.~\ref{fig:RKER} (d)) at which the ionization rate is dramatically enhanced \cite{zuo1995,peng2003,vafa2004,sabz2005}. Therefore, D$^{+}_{2}$ is ionized considerably but mainly through the low-energy band of the KER spectrum (Fig.~\ref{fig:tKER}). 
During 8 to 12.5 cycles, a considerable population of D$^{+}_{2}$, before reaching to the larger critical internuclear distances, is ionized and also the laser field becomes weak (Fig. 1,4, and 5). Therefore, after this time interval, the ionization channel for the negligible residual population of D$^{+}_{2}$ becomes nearly closed and the remaining D$^{+}_{2}$ ions will dissociate at longer times. 

%
\begin{acknowledgments}
We have benefited from valuable and stimulating discussions with Prof. H. Sabzyan, Dr. A. Shayesteh. We also thank Dr. I. V. Litvinyuk for the experimental data and his invaluable comments. We wish to acknowledge to thank the Shahid Beheshti University for the financial supports and research facilities and Isfahan High Performance Computing Center (IHPCC) and also Computational Nanotechnology Supercomputing Centre, Institute for Research in fundamental Sciences (IPM), P.O.Box 19395-5531, Tehran, Iran.
\end{acknowledgments}
%
\bibliography{p7}

\begin{thebibliography}{100}

\bibitem{review} 
F. Grossmann, Theoretical Femtosecond Physics: Atoms and Molecules in Strong Laser Fields (Springer-Verlag, 2008);
J. H. Posthumus, Rep. Prog. Phys, {\bf 67} 623 (2004);
A. Giusti-Suzor, F. H. Mies, L. F. DiMauro, E. Charron and B. Yang, J. Phys. B, {\bf 28}, 309 (1995).


\bibitem{Alnaser}
 A. S. Alnaser et al, Phys. Rev. Lett. {\bf 93}, 183202 (2004);

\bibitem{Litvin}  I. V. Litvinyuk and A. S. Alnaser and D. Comtois and D. Ray and A. T. Hasan and J-C kieffer and D. M. Villeneuve, New J. Phys., {\bf 10}, 083011 (2008).

\bibitem{pavicic2005}  D. Pavicic, A. Kiess, T. W. H\"{a}nsch and H. Figger, Phys. Rev. Lett. {\bf 94}, 163002 (2005).

\bibitem{Ergler2005}  Th. Ergler, A. Rudenko, B. Feuerstein, K. Zrost, C. D. Schr\"{o}ter, R. Moshammer, and J. Ullrich Phys. Rev. Lett. {\bf 95}, 093001 (2005).

\bibitem{Rudenko2007}  A. Rudenko, V. L. B. de Jesus, Th. Ergler, K. Zrost, B. Feuerstein, C. D. Schr\"{o}ter, R. Moshammer, and J. Ullrich, Phys. Rev. Lett. {\bf 99}, 263003 (2007).

\bibitem{Staudte2007PRL}  A. Staudte, D. Pavicic, S. Chelkowski, D. Zeidler, M. Meckel, H. Niikura, M. Sch\"{o}ffler, S. Sch\"{o}ssler, B. Ulrich, P. P. Rajeev, Th. Weber, T. Jahnke, D. M. Villeneuve, A. D. Bandrauk, C. L. Cocke, P. B. Corkum, and R. Dörner, Phys. Rev. Lett. {\bf 98}, 073003 (2007).

\bibitem{Staudte2007PRA}  S. Chelkowski, A. D. Bandrauk, A. Staudte and P. B. Corkum, Phys. Rev. A {\bf 76}, 013405 (2007).

\bibitem{vafa2008}  M. Vafaee, Phys. Rev. A {\bf 78}, 023410 (2008).

\bibitem{Hu2009}  S. X. Hu and L. A. Collins and B. I. Schneider, Phys. Rev. A {\bf 80}, 023426 (2009).


\bibitem{Su-Eberly1991}  Q. Su and J. H. Eberly, Phys. Rev. A {\bf 44}, 5997 (1991).



\bibitem{Saugout2008} S. Saugout, E. Charron and C. Cornaggia, Phys. Rev. A {\bf 77}, 023404 (2008);


\bibitem{Hiskes}  Hiskes J. R., Phys. Rev. {\bf 122}, 1207 (1961).
\bibitem{ban-lu2000}  A. D. Bandrauk and H. Z. Lu, Phys. Rev. A {\bf 62}, 053406 (2000).
\bibitem{vafa2004}  M. Vafaee and H. Sabzyan, J. Phys. B {\bf 37}, 4143 (2004).

%
\bibitem{splitting}
 B. Forenberg, A Practical Guide to Pseudospectral Methods (Cambridge University Press, 1996);
 C. Pozrikidis, Numerical Computations in Science and Engineering (Oxford: Oxford Univ., 1998); A. D.    Bandrauk and H. Shen, J. Chem. Phys. {\bf 99}, 1185 (1993).

\bibitem{vafa2006}  M. Vafaee, H. Sabzyan, Z. Vafaee and A. Katanforoush, http://arxiv.org/abs/physics/0509072

\bibitem{Feuerstein2003JPB}  B. Feuerstein and U. Thumm, J. Phys. B {\bf 36}, 707 (2003).

\bibitem{Coulomb Explosion Imaging}
S. Chelkowski, P.B. Corkum and A.D. Bandrauk, Phys. Rev. Lett. {\bf 82}, 3416 (1999).

\bibitem{peng2003}  L.-Y. Peng, D. Dundas, J. F. McCann, K. T. Taylor and I. D. Williams, J. Phys. B {\bf 36}, L295 (2003).

\bibitem{zuo1995}  T. Zuo and A. D. Bandrauk, Phys. Rev. A {\bf 52}, R2511 (1995).

\bibitem{sabz2005}  H. Sabzyan and M. Vafaee, Phys. Rev. A {\bf 71}, 063404 (2005).

\bibitem{cloumbimaging}  S. Chelkowski, P.B. Corkum and A.D. Bandrauk, Phys. Rev. Lett. {\bf 82}, 3416 (1999), S. Chelkowski and A.D. Bandrauk, Phys. Rev. A, {\bf 65}, 023403 (2002), C.R. Courtney and L.J. Frasinski, Phys. Lett. A {\bf 318}, 30 (2003).

 \end{thebibliography}

\end{document}